\documentclass[slac_one]{revtex4}
\usepackage{graphicx}
\usepackage{fancyhdr,color}
\definecolor{Blue}{named}{Blue}
\definecolor{Red}{named}{Red}
\definecolor{Green}{named}{ForestGreen}
\definecolor{Black}{named}{Black}
\definecolor{Olive}{named}{OliveGreen}

\begin{document}

\pagestyle{empty}

\vspace*{-1cm}
\begin{flushright}
CERN-PH-TH/2005-141\\
TSL/ISV-2005-0293\\
WUE-ITP-2005-008\\
hep-ph/0508313
\end{flushright}
                                                                                
\vspace*{1.4cm}
                                                                                
\begin{center}
                                                                                
{\Large
                                                                                
{\bf Identifying the NMSSM by the interplay of LHC and ILC\footnote{Contribution to the proceedings of the `2005 
International Linear Collider Workshop - Stanford, U.S.A.'}}}
                                                                                
\vspace{2cm}
{\large G.~Moortgat-Pick$^{1}$, S.~Hesselbach$^{2}$, F. Franke$^{3}$, H. Fraas$^{3}$}
\vspace{1cm}
                                                                                
{\it \noindent
$^{1}$~TH Division, Physics Department, CERN, CH-1211 Geneva 23, Switzerland\\
$^{2}$~High Energy Physics, Uppsala University, Box 535, S-75121 Uppsala, Sweden\\
$^{3}$~Institut f\"ur Theoretische Physik und Astrophysik, Universit\"at W\"urzburg,\\
}
                                                                                
\vspace{1cm}
                                                                                
\begin{minipage}{13cm}
The interplay between the LHC and the
$e^+ e^-$ International Linear Collider (ILC) with $\sqrt{s}=500$~GeV
might be crucial for the discrimination between the minimal and
next-to-minimal supersymmetric standard model.
We present an NMSSM scenario, where the light neutralinos have a significant singlino component,
that cannot be distinguished from the MSSM
by cross sections and mass measurements.
Mass and mixing state predictions for the heavier neutralinos from the ILC analysis at different
energy stages and
comparison with observation at the LHC, lead to clear identification of the particle character
and identify the underlying supersymmetric model.
\end{minipage}
\end{center}
                                                                                
\newpage

\title{{\small{Proceedings of '2005 International Linear Collider 
Workshop - Stanford,
U.S.A.'}}\\ 
\vspace{12pt}
Identifying the NMSSM by the interplay of LHC and ILC} 

%

\author{G. Moortgat-Pick}
\affiliation{TH Division, Physics Department, CERN, CH-1211 Geneva 23, 
Switzerland}
\author{S. Hesselbach}
\affiliation{High Energy Physics, Uppsala University, Box 535, 
S-751 21 Uppsala, Sweden}
\author{F. Franke, H. Fraas}
\affiliation{Institut f\"ur Theoretische Physik und Astrophysik, 
Universit\"at W\"urzburg,
        D-97074~W\"urzburg, Germany}


\begin{abstract}
The interplay between the LHC and the
$e^+ e^-$ International Linear Collider (ILC) with $\sqrt{s}=500$~GeV
might be crucial for the discrimination between the minimal and
next-to-minimal supersymmetric standard model. 
We present an NMSSM scenario, where the light neutralinos have a significant singlino component,
that cannot be distinguished from the MSSM 
by cross sections and mass measurements.
Mass and mixing state predictions for the heavier neutralinos from the ILC analysis at different 
energy stages and
comparison with observation at the LHC, lead to clear identification of the particle character
and identify the underlying supersymmetric model. 
\end{abstract}

\maketitle

\thispagestyle{fancy}


\section{INTRODUCTION} 

Supersymmetry (SUSY) is one of the most promising 
extensions to the Standard Model (SM).  Since low-energy SUSY is 
broken there exist numerous free parameters that make it a
highly challenging task to reveal the underlying model 
at the Large Hadron Collider (LHC) and at the International Linear Collider
(ILC).  It is planned that the ILC starts with an energy of
$\sqrt{s}=500$~GeV, which will be upgraded to about 1~TeV
\cite{ITRP}. However, already at the first energy stage, the ILC could
reach higher energy up to about $\sqrt{s}=650$~GeV at cost of
luminosity.  In this study we sketch a possible motivation to
apply this higher energy option.  Particularly interesting are case
studies which apply interplay of search strategies at the LHC and the
ILC~\cite{lhclc-report}. We extend in this paper the methods
for combined LHC/ILC analyses developed 
in~\cite{Moortgat-Pick:2005vs}.

An interesting possibility for the determination of the supersymmetric
model is to study the gaugino/higgsino particles, which are expected
to be among the lightest supersymmetric particles. In this paper we
consider two basic supersymmetric models: the minimal supersymmetric
standard model (MSSM) and the next-to-minimal supersymmetric standard
model (NMSSM). The MSSM contains four neutralinos $\tilde{\chi}^0_i$,
the mass eigenstates of the photino, zino and neutral higgsinos, and
two charginos $\tilde{\chi}^\pm_i$, being mixtures of wino and charged
higgsino. The neutralino/chargino sector depends at tree level on four
parameters: the U(1) and SU(2) gaugino masses $M_1$ and $M_2$, the
higgsino mass parameter $\mu$, and the ratio $\tan\beta$ of the vacuum
expectation values of the Higgs fields.  For the determination of
these parameters, straightforward strategies \cite{parameters,choi}
have been worked out even if only the light neutralinos and charginos
$\tilde{\chi}^0_1$, $\tilde{\chi}^0_2$ and $\tilde{\chi}^\pm_1$ are
kinematically accessible at the first stage of the ILC~\cite{ckmz}.

The NMSSM \cite{nmssm}
is the simplest extension of the MSSM
by an additional Higgs singlet field.
New parameters in the neutralino sector are the vacuum expectation
value $x$ of the singlet field and
the trilinear couplings $\lambda$ and $\kappa$ in the superpotential,
where the product $\lambda x = \mu_{\rm eff}$ replaces the
$\mu$-parameter of the MSSM \cite{Franke,Choi:2004zx}.
The additional fifth neutralino may significantly change the phenomenology
of the neutralino sector.
In scenarios where the lightest supersymmetric particle is a nearly
pure singlino, the existence of displaced vertices leads
to a particularly interesting experimental signature
\cite{singlinohugonie,Singlinos}. In case only a
part of the particle spectrum is kinematically accessible the
distinction between the models may become challenging. 

It has already been worked out that there exist
MSSM and NMSSM scenarios with
the same mass spectra of the light neutralinos but
different neutralino mixing. In this case
beam polarization is crucial for distinguishing the two models
\cite{Sitges}.
We present a scenario where
all kinematically accessible neutralinos and charginos have
similar masses and almost identical cross sections,
within experimental errors, in MSSM and NMSSM.
Although the second lightest neutralino in the NMSSM has a significant
singlino component, the models cannot be distinguished
by the experimental results at the LHC or at
the ILC$_{500}$ with $\sqrt{s}=500$ GeV alone 
if only measurements of masses, cross sections and
gaugino branching ratios are considered. 
Precision measurements of the neutralino branching ratio 
into the lightest Higgs particle and of  
the mass difference between the lightest and next-to-lightest 
SUSY particle \cite{gunion-mrenna}
may give first evidence for the SUSY model but are  
difficult to realize in our case. 
Therefore the identification of the
underlying model requires precision measurements of the
heavier neutralinos by combined analyses of LHC and ILC
as described in the following section.

\section{CASE STUDY}


We study an NMSSM scenario with the parameters 
\begin{eqnarray}
&&M_1=360~\mbox{\rm GeV},\quad M_2=147~\mbox{\rm GeV},\quad \tan\beta=10,\quad
\lambda=0.5,\quad x=915~\mbox{\rm GeV},\quad \kappa=0.2.
\label{eq-para-nm}
\end{eqnarray}
The hierarchy $M_1 > M_2$ of the U(1) and SU(2) mass parameters
leads to very similar masses
of the lightest neutralino $\tilde{\chi}_1^0$, which is assumed to be the
lightest supersymmetric particle (LSP), and of the light chargino
$\tilde{\chi}_1^\pm$. This mass degeneration 
is also typical for minimal anomaly mediated SUSY breaking
(mAMSB) scenarios.
Rather small mass differences may be resolved experimentally 
by applying the ISR method \cite{Hensel} at the linear 
collider \cite{gunion-mrenna} as well as at the LHC \cite{amsb-lhc}.

The NMSSM parameters lead to the following  gaugino/higgsino masses
and eigenstates: 
\begin{eqnarray}
m_{\tilde{\chi}^0_1}=138~\mbox{\rm GeV}, \quad\quad 
\tilde{\chi}^0_1=(-0.02, +0.97, -0.20, +0.09, -0.07), \label{ez-chi01-nm} \\
m_{\tilde{\chi}^0_2}=337~\mbox{\rm GeV},\quad\quad
\tilde{\chi}^0_2=(+0.62, +0.14, +0.25, -0.31, +0.65), \label{ez-chi02-nm} \\
m_{\tilde{\chi}^0_3}=367~\mbox{\rm GeV}, \quad\quad
\tilde{\chi}^0_3=(-0.75, +0.04, +0.01, -0.12, +0.65), \label{ez-chi03-nm} \\
m_{\tilde{\chi}^0_4}=468~\mbox{\rm GeV}, \quad\quad
\tilde{\chi}^0_4=(-0.03, +0.08, +0.70, +0.70, +0.08), \label{ez-chi04-nm} \\
m_{\tilde{\chi}^0_5}=499~\mbox{\rm GeV}, \quad\quad
\tilde{\chi}^0_5=(+0.21, -0.16, -0.64, +0.62, +0.37), \label{ez-chi05-nm}
\end{eqnarray}
where the neutralino eigenstates are given in the basis
$(\tilde{B}^0, \tilde{W}^0, \tilde{H}^0_1,\tilde{H}^0_2, \tilde{S})$.
As can be seen from eqs.~(\ref{ez-chi02-nm}) and (\ref{ez-chi03-nm}), 
the particles $\tilde{\chi}^0_2$ and $\tilde{\chi}^0_3$ have a rather 
strong singlino admixture.

The Higgs sector does not allow the identification of the NMSSM
\cite{NMSSMhiggs} if scalar and pseudoscalar Higgs bosons with
dominant singlet character escape detection. A scan with NMHDECAY
\cite{Ellwanger:2004xm} in our scenario over the remaining parameters
in the Higgs sector, $A_\lambda$ and $A_\kappa$, results in parameter
points which survive the theoretical and experimental constraints in
the region $2740~\textrm{GeV} < A_\lambda < 5465$~GeV and
$-553~\textrm{GeV} < A_\kappa < 0$. For $-443~\textrm{GeV} < A_\kappa
< -91$~GeV the second lightest scalar ($S_2$) and the lightest
pseudoscalar ($P_1$) Higgs particle have very pure singlet character
and are heavier than the mass difference
$m_{\tilde{\chi}^0_3} - m_{\tilde{\chi}^0_1}$,
hence the decays of the neutralinos $\tilde{\chi}^0_2$ and
$\tilde{\chi}^0_3$, which will be discussed in the following, are not
affected by $S_2$ and $P_1$, see figure~\ref{higgs-sector} (left panel). For
our specific case study we choose $A_{\lambda}=4000$~GeV and
$A_{\kappa}=-200$~GeV, which leads to $m_{S_2}=311$~GeV,
$m_{P_1}=335$~GeV and $m_{S_3}$, $m_{P_2}$ and $m_{H^{\pm}}>4$~TeV.
Furthermore the lightest scalar Higgs $S_1$ has MSSM-like character in
this parameter range with a mass of about 124~GeV.  Also the branching
ratio of $\tilde{\chi}^0_2$ in the lightest Higgs particle differs
only by a factor two in both scenarios. In case that a precise
measurement of this BR is possible first hints for the
inconsistency of the model could be derived at the ILC.

\subsection{Strategy for the gaugino/higgsino sector}
In our NMSSM scenario in the gaugino/higgsino sector only the
light chargino $\tilde{\chi}^{\pm}_1$  and the light neutralinos
$\tilde{\chi}_1^0$ and
$\tilde{\chi}_2^0$
are accessible at the ILC$_{500}$.
We calculate the masses of the charginos and neutralinos and
the cross sections for the pair production
of the light chargino $e^+e^- \rightarrow \tilde{\chi}^+_1
\tilde{\chi}^-_1$ and for the associated production of the light neutralinos
$e^+e^- \rightarrow \tilde{\chi}_1^0 \tilde{\chi}_2^0$ with 
polarized and unpolarized beams.
 
\begin{figure*}[t]
\setlength{\unitlength}{1cm}
\begin{center}
\begin{picture}(10,5)
\put(-3.4,4.55){\small $m_{H}/$GeV}
\put(-.5,4.1){\color{Blue}\small $P_1$}
\put(-.5,1.8){\color{Red}\small $S_2$}
\put(-.5,.8){\color{Green}\small $S_1$}
\put(0,-.5){\small $A_{\kappa}$/GeV}
\put(-3,-1){\includegraphics[width=70mm]{0206-higgsmass.epsy}}
\put(8.7,3.5){\small{\color{Red} \phantom{$\longleftarrow$ }$\tilde{\chi}^0_3$\quad} and
{\color{Blue}\quad $\tilde{\chi}^0_4$}}
\put(4.9,4.6){\small $m_{\tilde{\chi}^0_i}$/GeV}
\put(4.5,1.2){\small LHC:}
\put(4.5,.9){\small measurement}
\put(4.95,.6){\small of \color{green}$m_{\tilde{\chi}^0_i}\rightarrow$}
\put(7.5,5){\small Contradiction within MSSM}
\put(8.8,4.5){\small ILC: prediction of}
\put(8.8,4){\small  mixing character of}
\put(8.3,-.5){\small gaugino character}
\put(6,-1){\includegraphics[width=70mm]{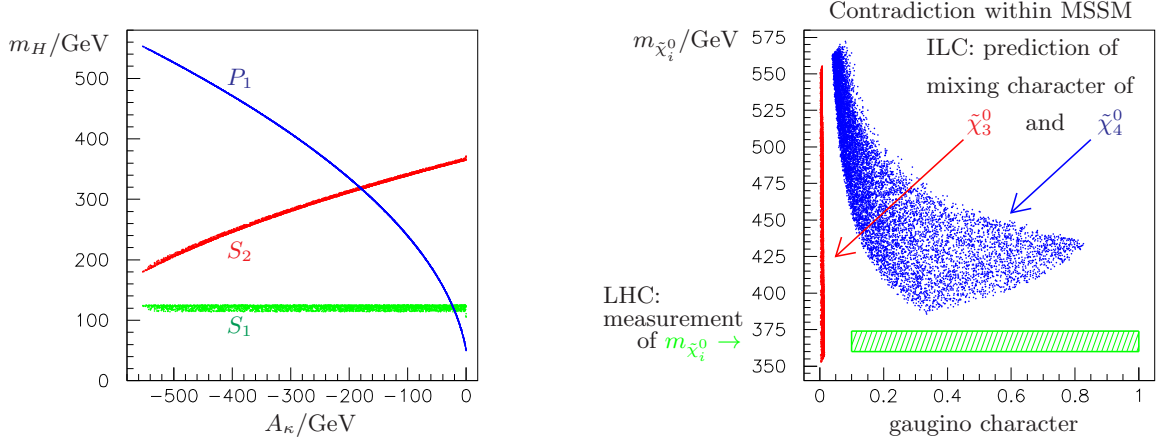}}
\end{picture}
\end{center}
\vspace{-.2cm}
\caption{Left: The possible
masses of the two light scalar Higgs bosons, $m_{S_1}$, $m_{S_2}$, and of the
lightest pseudoscalar Higgs boson $m_{P_1}$ as function of the trilinear
Higgs parameter $A_{\kappa}$ in the NMSSM.
In our chosen scenario, $S_1$ is MSSM-like and
$S_2$ and $P_1$ are heavy singlet-dominated Higgs particles.
Right: Predicted masses and gaugino admixture for the heavier neutralinos
$\tilde{\chi}^0_3$  and $\tilde{\chi}^0_4$ within the consistent
parameter ranges derived at the ILC$_{500}$ analysis
in the MSSM and measured mass $m_{\tilde{\chi}^0_i}=367\pm 7$~GeV
of a neutralino with sufficiently high gaugino admixture in cascade
decays at the LHC. We took a lower bound of sufficient gaugino admixture
of about 10\% for
the heavy neutralinos, cf.\ \cite{Giacomo}. \label{higgs-sector}}
\end{figure*}

The masses and cross sections in different beam polarization configurations
provide the experimental input for deriving the supersymmetric parameters 
within the MSSM using standard methods
\cite{choi,ckmz}:
\begin{itemize}
\item We assume an uncertainty of $\mathcal{O}(1-2\%)$ for the masses
$m_{\tilde{\chi}^{\pm}_1}$,
$m_{\tilde{\chi}^0_1}$, $m_{\tilde{\chi}^0_2}$,
$m_{\tilde{\nu}_e}$, $m_{\tilde{e}_L}$ and $m_{\tilde{e}_R}$.
The errors of the cross sections,
$e^+e^- \rightarrow \tilde{\chi}^+_1 \tilde{\chi}^-_1$ and
$e^+e^- \rightarrow \tilde{\chi}_1^0 \tilde{\chi}_2^0$, are composed
of the error due to the mass uncertainties, polarization uncertainty and
one standard deviation statistical error based on
$\int {\cal L}=100$~fb$^{-1}$ for each polarization configuration.
Deviations in the cross sections due to 
the polarization uncertainty of 
$\Delta P_{e^{\pm}}/P_{e^{\pm}}=0.5\%$ are generally small; it 
is expected that the error
could even be reduced up to  $\Delta P_{e^{\pm}}/P_{e^{\pm}}=0.2\%$--$0.1\%$,
\cite{Power}. The assumed uncertainties in total are listed in 
table~\ref{lc-input}.
\item
From the chargino mass $m_{\tilde{\chi}^{\pm}_1}$ and
the cross section
$e^+e^- \rightarrow \tilde{\chi}^+_1 \tilde{\chi}^-_1$
measured at two energies, $\sqrt{s}=400$~GeV and $500$~GeV,
we determine bounds for the
elements $U_{11}$ and $V_{11}$ of the chargino mixing matrices:
\begin{equation}
U^2_{11}=[0.84,1.0], \quad V^2_{11}=[0.83,1.0].
\end{equation}
Polarized beams allow the resolution of ambiguities and the improvement
of the accuracy.
\item Using the mixing matrix elements $U_{11}$ and $V_{11}$,
the masses $m_{\tilde{\chi}^{\pm}_1}$,
$m_{\tilde{\chi}^0_1}$ and $m_{\tilde{\chi}^0_2}$,
and the cross sections for
$e^+e^- \rightarrow \tilde{\chi}^0_1 \tilde{\chi}^0_2$,
we derive constraints for
the parameters $M_1$, $M_2$, $\mu$ and $\tan\beta$:
\begin{eqnarray} 
M_1 &=& 377 \pm 42 \mbox{ GeV}, \label{MSSMresult1}\\
M_2 &=& 150\pm 20 \mbox{ GeV}, \\
\mu &=& 450 \pm 100 \mbox{ GeV}, \\
\tan\beta &=& [1,30]. \label{MSSMresult4}
\end{eqnarray}
Note that, in our scenario with $M_1 > M_2$, the crucial observable to
determine the parameter $M_1$ is $m_{\tilde{\chi}^0_2}$ and not
$m_{\tilde{\chi}^0_1}$ as often assumed. Such a hierarchy could be
naturally embedded in mAMSB scenarios. For even larger $M_1\gg M_2$
the heavier neutralinos $\tilde{\chi}^0_{3,4}$ become crucial for $M_1$
determination see \cite{Moortgat-Pick:2005vs,m1-paper}.  
Since the  heavier  
neutralino and chargino states are not produced, some of the 
parameters ---in our case $\mu$ and $\tan\beta$--- 
can only be determined with a considerable
uncertainty.

Within these limits an explicit MSSM scenario, 
\begin{eqnarray}
&&M_1=375~\mbox{\rm GeV},\quad M_2=152~\mbox{\rm GeV},\quad \tan\beta=8,\quad
\mu=360~\mbox{\rm GeV},
\label{eq-para-ms}
\end{eqnarray}
leads to the 
same (lighter) neutralino/chargino masses and cross sections:
\begin{eqnarray}
m_{\tilde{\chi}^0_1}=138~\mbox{\rm GeV}, \quad\quad 
\tilde{\chi}^0_1=(+0.03, -0.96, +0.26, -0.13), \label{ez-chi01-ms}\\
m_{\tilde{\chi}^0_2}=344~\mbox{\rm GeV}, 
\quad\quad
\tilde{\chi}^0_2=(+0.72, +0.22, +0.48, -0.46), \label{ez-chi02-ms} \\
m_{\tilde{\chi}^0_3}=366~\mbox{\rm GeV}, \quad\quad
\tilde{\chi}^0_3= (-0.04, +0.10, -0.70, -0.71), \label{ez-chi03-ms} \\
m_{\tilde{\chi}^0_4}=410~\mbox{\rm GeV}, \quad\quad
\tilde{\chi}^0_4=(-0.70, +0.18, +0.47, -0.52), \label{ez-chi04-ms}
\end{eqnarray}
where the neutralino mixing states are given in the basis
$(\tilde{B}^0, \tilde{W}^0, \tilde{H}^0_1,\tilde{H}^0_2)$.
Comparing eqs.~(\ref{ez-chi01-ms})--(\ref{ez-chi03-ms}) with 
eqs.~(\ref{ez-chi01-nm})--(\ref{ez-chi03-nm}) shows that 
the three lightest neutralino masses are the same within the 
experimental uncertainties. We checked that also the accessible
cross sections at the ILC$_{500}$ 
and the BR's of $\tilde{\chi}^0_2$ are consistent.

\item After the determination of the fundamental MSSM
parameters we calculate the heavy chargino
and neutralino masses and expected mixing characters.
For the masses we obtain:
\begin{equation}
m_{\tilde{\chi}^0_3} = 443\pm 107 \mbox{ GeV},\quad\quad 
m_{\tilde{\chi}^0_4} = 490\pm 110 \mbox{ GeV}, \quad\quad
m_{\tilde{\chi}^{\pm}_2} = 475 \pm 125 \mbox{ GeV}. \label{eq_mp3_nm}
\end{equation}
\end{itemize}
The predicted gaugino admixture of $\tilde{\chi}^0_3$, $\tilde{\chi}^0_4$
within the allowed parameter ranges, eqs.(\ref{MSSMresult1})--(\ref{MSSMresult4}),
are shown in figure~\ref{higgs-sector} (right panel). Obviously, the heavy neutralino
$\tilde{\chi}^0_3$ should be almost a pure higgsino within the MSSM prediction. 
The predicted properties of the
heavier particles can now be compared with 
mass measurements of such SUSY particles via the analysis
of cascade decays at the LHC~\cite{lhclc-report}.

We emphasize that although we started with an NMSSM scenario where
$\tilde{\chi}^0_2$ and $\tilde{\chi}^0_3$ have large singlino
admixtures, the MSSM parameter strategy does not fail and
the experimental results from the ILC$_{500}$ with $\sqrt{s}=400$~GeV
and $500$~GeV lead to a consistent parameter determination in the MSSM.
Hence in the considered scenario the analyses at the ILC$_{500}$ or
LHC alone do not allow a clear discrimination between MSSM and NMSSM.
All predictions for the heavier gaugino/higgsino masses are
consistent with both models.
However, the ILC$_{500}$ analysis predicts an almost pure
higgsino-like state for $\tilde{\chi}^0_3$
and a mixed gaugino-higgsino-like $\tilde{\chi}^0_4$, see
figure~\ref{higgs-sector} (right panel). 
This allows the identification of the underlying supersymmetric model
in combined analyses at the LHC and the
ILC$_{650}^{{\cal L}=1/3}$.

\subsection{Interplay between LHC and ILC}
The expected large cross sections for squark and gluino
production at the LHC
give access to a large spectrum of coloured as
well as non-coloured supersymmetric particles via the cascade decays.
Heavy gaugino-states appear almost only in cascade decays and 
there exist some true simulations how to measure the heavier gauginos in
such decays at the LHC \cite{Giacomo}. Particularly helpful for the
identification 
of the particles involved in the cascades, e.g.\ for more
model-independent analyses,  
are  mass predictions from the ILC analysis which lead to an increase
of statistical  
sensitivity for the LHC analysis and open the possibility of identifying even
marginal signals in the squark cascades~\cite{lhclc-report}. 
However, since higgsino-like charginos and neutralinos
do not couple to squarks, their detection via cascade decays is not possible.

In our original NMSSM scenario 
the neutralinos $\tilde{\chi}^0_2$ and
$\tilde{\chi}^0_3$ have a large
bino-admixture and therefore appear
in the squark decay cascades. The dominant decay mode of
$\tilde{\chi}^0_2$ has a branching ratio $BR(\tilde{\chi}^0_2\to
\tilde{\chi}^{\pm}_1 W^{\mp})\sim 50\%$, while
for the $\tilde{\chi}^0_3$ decays $BR(\tilde{\chi}^0_3\to
\tilde{\ell}^{\pm}_{L,R} \ell^{\mp})\sim 45\%$ is largest.
Since the heavier neutralinos, $\tilde{\chi}^0_4$, $\tilde{\chi}^0_5$,
are mainly higgsino-like,
no visible edges from these particles occur in the cascades.
It is expected to see the edges for
$\tilde{\chi}^0_2 \to \tilde{\ell}^{\pm}_R \ell^{\mp}$,
$\tilde{\chi}^0_2 \to \tilde{\ell}^{\pm}_L \ell^{\mp}$,
$\tilde{\chi}^0_3 \to \tilde{\ell}^{\pm}_R \ell^{\mp}$
and for
$\tilde{\chi}^0_3 \to \tilde{\ell}^{\pm}_L \ell^{\mp}$
\cite{Giacomo_private}.

With a precise
mass measurement of
$\tilde{\chi}^0_1$,$\tilde{\chi}^0_2$, $\tilde{\ell}_{L,R}$ and
$\tilde{\nu}$ from the ILC$_{500}$
analysis, a clear identification and separation of
the edges of the two gauginos at the LHC is
possible without imposing specific model assumptions.
We therefore assume a precision of
about 2\% for the measurement of $m_{\tilde{\chi}^0_3}$, in analogy to
\cite{Giacomo}:
\begin{eqnarray} 
\label{eq:mchi3LHC}
&& m_{\tilde{\chi}^0_3}=
367\pm 7 \mbox{ GeV}.
\end{eqnarray}
The precise mass measurement of $\tilde{\chi}^0_3$ is compatible with
the mass predictions of the ILC$_{500}$ but not with the prediction of
the mixing character,
see e.g.\ eq.~(\ref{ez-chi03-ms}). 
However, it is not clear that
the measured particle at the LHC is indeed the $\tilde{\chi}^0_3$. Often
in the constrained MSSM, as e.g. also in our MSSM comparison scenario,
the second heaviest neutralino $\tilde{\chi}^0_3$ is nearly a pure
higgsino and does not couple in the cascade decays. In those cases,
the heaviest neutralino $\tilde{\chi}^0_4$ has frequently a sufficiently
large gaugino component and could be measured in cascades, as shown in
\cite{lhclc-report,Giacomo}.

Therefore is is inevitable to  
discuss the following cases of possible particle {\it identification}
of the measured gaugino mass $m_{\tilde{\chi}^0_3}$ at the LHC:
\begin{itemize}
\item interpretation of the measured particle as
$\tilde{\chi}^0_3$ and feeding it back in the ILC analysis leads to
improved parameter determination and mass prediction for
$m_{\tilde{\chi}^0_4}$, $m_{\tilde{\chi}^{\pm}_2}$.
Using eq.~(\ref{eq:mchi3LHC}) for the ILC$_{500}$ analysis
leads in our case, after rechecking with the allowed cross sections of
$\tilde{\chi}^0_1\tilde{\chi}^0_2$ and
$\tilde{\chi}^{+}_1\tilde{\chi}^-_1$ production, 
to rather precise mass predictions:
\begin{equation}
m_{\tilde{\chi}^0_4}=[384, 393]\mbox{ GeV \quad and \quad}
m_{\tilde{\chi}^{\pm}_2}=[360, 380]\mbox{ GeV}.
\label{eq:heavymass-pred}
\end{equation}
\item interpretation of the measured particle as $\tilde{\chi}^0_4$
and feeding it back in the parameter determination of the ILC analysis
leads to inconsistency with the measured cross sections of
$\tilde{\chi}^0_1\tilde{\chi}^0_2$ and $\tilde{\chi}^{+}_1\tilde{\chi}^-_1$ production.
\end{itemize}
The combined LHC$\leftrightarrow$ILC$_{500}$ analysis leads therefore
to a correct interpretation of the measured particles in the cascades.
However, a neutralino $\tilde{\chi}^0_3$  with sufficiently large
gaugino admixture to couple to squarks is incompatible with the
allowed parameter ranges of
eqs.~(\ref{MSSMresult1})--(\ref{MSSMresult4}) in the MSSM, 
cf.\ figure~\ref{higgs-sector} (right panel).

We point out
that a measurement of the neutralino masses
$m_{\tilde{\chi}^0_1}$, $m_{\tilde{\chi}^0_2}$, $m_{\tilde{\chi}^0_3}$
which could take place at the LHC alone is not sufficient to distinguish
the SUSY models since rather similar mass spectra could exist, cf. 
eqs.~(\ref{ez-chi01-nm})--(\ref{ez-chi03-nm}) 
with eqs.~(\ref{ez-chi01-ms})--(\ref{ez-chi03-ms}).

Therefore the cross sections in different beam polarization
configurations
at the ILC have to be included in the analysis. 
The combined results from the LHC and the ILC$_{500}$ analyses
and the rather precise predictions for the missing
chargino/neutralino masses, eq.~(\ref{eq:heavymass-pred}),
constitute a serious motivation
to apply immediately the low-luminosity but higher-energy option
ILC$_{650}^{{\cal L}=1/3}$, 
which finally leads to the right identification
of the underlying model. The expected polarized and
unpolarized cross sections, including the statistical error on the basis of
one third of the luminosity of the ILC$_{500}$, are given in
Table~\ref{tab_heavy}. The neutralino
$\tilde{\chi}^0_3$ as well as the higgsino-like heavy neutralino
$\tilde{\chi}^0_4$ and the chargino $\tilde{\chi}^{\pm}_2$ are now
accessible at the ILC$^{{\cal L}=1/3}_{650}$. 
Already the high rates for $\tilde{\chi}^0_1\tilde{\chi}^0_3$ production
give last true evidence for the obvious contraction with an
corresponding MSSM scenario.
Together with the 
mass measurements of $m_{\tilde{\chi}^0_4}=468$~GeV and 
$m_{\tilde{\chi}^{\pm}_2}=474$~GeV, 
which are also in strong disagreement with the mass prediction, 
eq.~(\ref{eq:heavymass-pred}), one has sufficient observables which point
to the NMSSM. Extensions of existing fit programs for the NMSSM
may lead to an exact resolution of the 
underlying parameters~\cite{fit-porod}.

\section{CONCLUSIONS}
We have presented a scenario in the next-to-minimal supersymmetric
standard model
(NMSSM) that could not be distinguished from the MSSM at either the LHC or at 
the first stage of the International Linear Collider with
$\sqrt{s}= 500$ GeV. It turns out that the most promising sector for distinction 
is the gaugino/higgsino sector.
Although a light neutralino has a significant
singlet component in the NMSSM, the masses of the accessible
light neutralinos and charginos, as well as the production cross sections,
lead to identical values in the two models within experimental errors.
The comparison of the predicted masses and mixing character of the heavier neutralinos and charginos
with the measured masses 
in combined analyses with the LHC followed
by a precise measurement of the cross sections at the ILC at
$\sqrt{s}= 650$ GeV leads to a clear identification of the supersymmetric
model.

The exemplary scenario shows that the interplay between the two
experiments could be crucial for the determination of the supersymmetric
model. A possible feed-back of ILC$_{500}$/LHC results 
could motivate the immediate use of the low-luminosity option of
the ILC  at $\sqrt{s}=650$~GeV in order to resolve model ambiguities even at an early stage of
the experiment and outline future search strategies at the
upgraded ILC at 1 TeV.

\begin{table}[t!]
\renewcommand{\arraystretch}{1.3}
\centering
\begin{tabular}{|c||c|c|c|c|}
\hline
$m_{\tilde{\chi}^0_1}$/GeV$=138\pm 2.8$  &
 & \multicolumn{2}{|c|}{$\sigma(e^+e^-\to\tilde{\chi}^{\pm}_1\tilde{\chi}^{\mp}_1)$/fb} & $\sigma(e^+e^-\to\tilde{\chi}^{0}_1\tilde{\chi}^{0}_2)$/fb \\
$m_{\tilde{\chi}^0_2}$/GeV$=337\pm 5.1$ &  $(P_{e^-},P_{e^+})$
& $\sqrt{s}=400$~GeV & $\sqrt{s}=500$~GeV & $\sqrt{s}=500$~GeV  \\
\cline{2-5}
$m_{\tilde{\chi}^{\pm}_1}$/GeV$=139\pm 2.8$ &
Unpolarized & $323.9 \pm 33.5$ & $287.5 \pm 16.5$ & $4.0 \pm 1.2$\\
$m_{\tilde{e}_L}$/GeV$=240\pm 3.6$ &  $(-90\%,+60\%)$ &
 $984.0 \pm 101.6$ & $873.9 \pm 50.1$ & $12.1 \pm 3.8$ \\
$m_{\tilde{e}_R}$/GeV$=220\pm 3.3$ &   $(+90\%,-60\%)$ &
$13.6 \pm 1.6$ & $11.7 \pm 1.2$ & $0.2 \pm 0.1$\\
$m_{\tilde{\nu}_e}$/GeV$=226\pm 3.4$ & &&&\\ \hline
\end{tabular}
\caption{Masses with 
1.5\% ($\tilde{\chi}^0_{2,3}$, $\tilde{e}_{L,R}$, $\tilde{\nu}_e$)
and 2\% ($\tilde{\chi}^0_1$, $\tilde{\chi}^{\pm}_1$)
uncertainty and cross sections with an error
composed of
the error due to the mass uncertainties,  polarization uncertainty and
one standard deviation statistical error based on
$\int {\cal L}=100$~fb$^{-1}$,
for both unpolarized beams and polarized beams with
$(P_{e^-},P_{e^+})=(\mp 90\%,\pm 60\%)$ and $\Delta P(e^\pm)/P(e^{\pm})=0.5\%$,
in analogy to the study in \cite{lhclc-report}.
\label{lc-input}
}
\end{table}

\begin{table}
\renewcommand{\arraystretch}{1.2}
\centering
\begin{tabular}{|c|c|c|c|c|}
\hline  &
 \multicolumn{3}{c|}{%
$\sigma(e^+e^- \to \tilde{\chi}^0_1\tilde{\chi}^0_j)/$fb at
$\sqrt{s}=650$~GeV} & $\sigma(e^+e^- \to \tilde{\chi}^{\pm}_1\tilde{\chi}^{\mp}_2)/$fb at
$\sqrt{s}=650$~GeV \\     \cline{2-4}
 & \makebox[21mm]{$j=3$} & \makebox[21mm]{$j=4$} & \makebox[21mm]{$j=5$} &\\
\hline \hline
Unpolarized beams & $12.2 \pm 0.6$
& $5.5 \pm 0.4$ & $\le 0.02$ &  $2.4 \pm 0.3$ \\ \hline
$P(e^-)=-90\%$, $P(e^+)=+60\%$ &
$36.9 \pm 1.1$ & $14.8 \pm 0.7$ & $\le 0.07$ & $5.8 \pm 0.4$ \\ \hline
$P(e^-)=+90\%$, $P(e^+)=-60\%$ &
$0.6 \pm 0.1$ & $2.2 \pm 0.3$ & $\le 0.01$ & $1.6 \pm 0.2$\\ \hline
\end{tabular}
\caption{Expected cross sections for
the associated production of
the heavier neutralinos and charginos in the NMSSM  scenario
for the
ILC$^{{\cal L}=1/3}_{650}$ option with one sigma
statistical error
based on $\int {\cal L} = 33$~fb$^{-1}$
for both unpolarized and polarized beams.
\label{tab_heavy}}
\end{table}

%



\begin{acknowledgments}
We are very grateful to G.~Polesello and P.~Richardson for
constructive discussions.
S.H.\ is supported by the G\"oran Gustafsson Foundation.
This work is supported by the
European Community's Human Potential Programme under contract
HPRN-CT-2000-00149 and by the Deutsche Forschungsgemeinschaft (DFG) under
contract No.\ \mbox{FR~1064/5-2}.
\end{acknowledgments}

\end{document}